\documentclass[12pt,preprint]{aastex}

\slugcomment{} \shorttitle{NIR Monitoring of UCDs}
\shortauthors{Blake et al.}

\begin{document}
\title{Near Infrared Monitoring of Ultracool Dwarfs:\\ Prospects for Searching for Transiting
Companions}

\author{Cullen H. Blake\altaffilmark{1}} \affil{Harvard-Smithsonian Center for
Astrophysics, Cambridge, MA 02138; cblake@cfa.harvard.edu}
\altaffiltext{1}{Harvard Origins of Life Initiative Fellow}

\author{Joshua S. Bloom\altaffilmark{2}} \affil{Astronomy Department,
University of California, Berkeley, CA 94720}
\altaffiltext{2}{Sloan Research Fellow}

\author{David W. Latham} \affil{Harvard-Smithsonian Center for
Astrophysics, Cambridge, MA 02138}

\author{Andrew H. Szentgyorgyi} \affil{Harvard-Smithsonian Center for
Astrophysics, Cambridge, MA 02138}

\author{Michael F. Skrutskie} \affil{Department of Astronomy, P.O. Box 3818, University of Virginia, Charlottesville, VA 22903-0818}

\author{Emilio E. Falco} \affil{Harvard-Smithsonian Center for
Astrophysics, Cambridge, MA 02138}

\author{Dan S. Starr} \affil{Astronomy Department,
University of California, Berkeley, CA 94720}
\affil{Las Cumbres Global Telescope Network, 6740 Cortona Dr., Santa Barbara, CA 93117}

\begin{abstract}
Stars of late-M and L spectral types, collectively known as Ultracool
Dwarfs (UCDs), may be excellent targets for searches for extrasolar
planets. Owing to their small radii, the signal from an Earth-size
planet transiting a UCD is, in principle, readily detectable. We
present results from a study designed to evaluate the feasibility of
using precise near infrared (NIR) photometry to detect terrestrial
extrasolar planets orbiting UCDs. We used the Peters Automated
InfRared Imaging TELescope (PAIRITEL) to observe a sample of 13 UCDs
over a period of 10 months. We consider several important systematic
effects in NIR differential photometry and develop techniques for
generating photometry with a precision of 0.01 mag and long-term
stability. We simulate the planet detection efficiency of an extended
campaign to monitor a large sample of UCDs with PAIRITEL. We find that
both a targeted campaign with a single telescope lasting several years
and a campaign making use of a network of telescopes distributed in
longitude could provide significant sensitivity to terrestrial planets
orbiting UCDs, potentially in the habitable zone.
\end{abstract}


\section{Introduction}
Theories of planet formation make different predictions about the
frequency, mass, and separations of planetary companions to stars of different
masses.  The existence of small ($M\ll M_{\rm{J}}$) companions to
small stars at separations $<10$ AU, and a relative paucity of large ($M\approx
M_{\rm{J}}$) companions, is a general prediction of the core accretion
model of planet formation \citep{Adams2005}. \citet{Ida2005} found
that planet formation via core accretion can be efficient around
low-mass stars and that it is even possible for small planets to open
a gap in the protoplanetary disk and move inwards through Type II
migration. Recently, \citet{Payne2007} studied the specific case of
planet formation around Ultracool Dwarfs (UCDs\footnote{Here we follow the convention of
\citet{Bailer-Jones2002} and define the term Ultracool Dwarf (UCD) to
include objects of spectral type later than about M7, including L and
T dwarfs. Such objects may be bonafide brown dwarfs or very small
hydrogen-burning stars.}) and found that the formation of terrestrial
planets up to 5 M$_{\earth}$ should be possible, though migration of
these planets inward may be unlikely. The formation of large
companions can also be explained within the disk instability
paradigm. \citet{Boss2006} recently suggested that searching for
massive companions to UCDs is an important way to differentiate
between these two competing planet formation scenarios. There are initial indications from radial-velocity searches and
microlensing surveys that early-M dwarfs have a relative paucity of
Jupiter-mass companions compared to Sun-like stars \citep{Butler2004,
Gaudi2002}. Recently, there has been evidence that lower mass
companions to M dwarf stars may be a common occurrence. The microlensing detections
of super-Earth and Neptune mass planets orbiting M dwarfs by \citet{Beaulieu2006} and \citet{Gould2006b} indicates that these types of planets may be relatively common.

There are more than 2000 UCDs known today. These objects, many discovered with the 2-Micron All Sky Survey (2MASS, \citealt{Cruz2003}), are a well-studied class spanning the M, L, and T spectral types.
Little is known about the formation of planets around UCDs. Disks around young UCDs, which would be  a starting point for all planet formation mechanisms, appear to be quite common. \citet{Luhman2005}  found that around 50\% of the UCDs in two star-forming regions show evidence for infrared excess attributed to the presence of a disk. This disk fraction is consistent with the disk fraction of larger stars also observed in those star forming regions. The authors argue that this is strong evidence for a common formation mechanism of stars and UCDs and that the raw materials for planet formation are available near UCDs as often as near early M-type stars. Observations of UCD disks by \citet{Apai2005} show that dust settling and grain growth are taking place in these disks, indicating that an important first step in planet formation is occurring in disks surrounding UCDs.

Current planet searches pay relatively little attention to stars of spectral type later than about M4. There are seven known M dwarf planetary systems (GJ 876: \citealt{Delfosse1998,Rivera2004}, GJ 436: \citealt{Butler2004}, GJ 581: \citealt{Bonfils2005}, GJ 674: \citealt{Bonfils2007}, GJ 317: \citealt{Johnson2007}, GJ 849: \citealt{Butler2006}, and GJ 176: \citealt{Endl2007})  from radial velocity searches, but a search for planetary companions to UCDs will likely require observations in the infrared. High-precision optical radial velocities of early-M dwarfs have already been reported in the literature \citep{Endl2003}, and the techniques for obtaining radial velocities of UCDs in the NIR are approaching the level required for confirming massive, short-period companions \citep{Blake2007a}. A transit search offers the exciting possibility of detecting close-in companions to UCDs and, because of the relatively small size of the host UCDs, enables, in principle, the detection of terrestrial planets.  The possibility of detecting terrestrial planetary companions to UCDs and M dwarfs through observations of transits has been explored by \citet{Gould2003}, \citet{Blake2003}, \citet{Caballero2003}, \citet{Snellen2005}, \citet{Plavchen2005,Plavchen2007}, and \citet{Nutzman2007}. If photometric precision similar to that realized by transit searches at optical wavelengths can be achieved in observations of UCDs, rocky companions as small as Earth would be detectable. 

Searching for transits of UCDs presents many challenges. These objects are intrinsically faint and therefore very rare in the shallow, wide-field surveys typically used to find transiting planets.  Most current transit searches exploit a multiplexing advantage by observing many thousands of stars simultaneously (i.e. \citealt{Bakos2004}), but the low density of UCDs on the sky requires observing them individually.  While photometric precision sufficient for the detection of transits is regularly achieved by surveys operating in the optical regime, the UCDs are much more luminous in the NIR than in the optical. Therefore, observations must be carried out in a wavelength regime where observational techniques are not as well developed and it may not be possible to design suitable wide-field instruments. Several authors have  discussed these challenges and the prospects for high precision, NIR, time-series photometry. In addition to the results presented by \citet{Koen2005} and \citet{Bailer-Jones2003}, it has been demonstrated by \citet{Snellen2005} that modern ground-based NIR detectors are capable of producing 0.1\% differential photometry of bright point sources under certain circumstances. 

It is important to ascertain the level of intrinsic variability of the UCDs themselves.  While large photometric surveys, such as microlensing searches and transiting planet surveys, have provided a wealth of information concerning the photometric properties of larger and bluer stars, relatively little is known about the time-dependent behavior of UCDs. For a transit search, we are primarily concerned with variability on timescales comparable to the duration of the transit events. Their complex, dynamical atmospheres, in conjunction with  their observed rapid rotation, could result in observable photometric variability due to spots. Many late M dwarfs are thought to have strong magnetic fields and are found to flare. These flares, which can be large in amplitude in the U band (i.e. \citealt{Rockenfeller2006}), are found to strongly decrease in amplitude with wavelength. There is little evidence in the literature for similar flares on L dwarfs. There is a growing body of literature concerning ``weather'' and the intrinsic variability of UCDs at both optical and NIR  wavelengths \citep{Gelino2002, Bailer-Jones2003,Koen2005}. Observations spanning the optical to the mid-infrared have produced conflicting accounts of the intrinsic variability of L dwarfs. While it seems clear that these objects are not found to be so variable as to preclude the detection of transiting planets, it is not clear at what level  they do exhibit variability. Recent work by \citet{Calderon2006} demonstrated that in the mid-infrared at least one L dwarf is photometrically stable at the level of 3 mmag.

Here, we present the results of a pilot study designed to evaluate the feasibility of a NIR, targeted search for transiting terrestrial companions to UCDs. In Section 2 we outline the basic properties of the types of planetary systems that we could hope to detect. In Section 3 we describe the robotic observing system used to gather these observations and the process of data reduction and photometry. In Section 4 we describe some of the problems unique to differential photometry of cool stars in the NIR. In Section 5 we summarize the overall quality of the NIR observations of UCDs and present the results of simulations designed to evaluate the feasibility of detecting terrestrial planets in with a survey like ours.

\section{UCD Planetary Systems}
A search for transiting planetary companions is most likely to detect objects orbiting close to their host. These planets will spend an
appreciable part of their orbit in transit and have orbital periods short enough so that the probability of detection is high during an
extended observing campaign. A practical limit to the minimum orbital radius of the companion is given by the Roche limit, $a_{\rm R}$. The relation of the observed distribution of close-in extrasolar planets to the Roche limit was explored by \citet{Ford2005} who determined that the distribution may extend inward to about twice the Roche limit as derived by \citet{Faber2005} and \citet{Paczynski1971}:

\begin{equation}
a_{R}=2.165 \times R_{p}  \left(M_{UCD}\over{M_{p}}\right)^{1/3} = 9.2\times10^{-5} \left(R_{p}\over{R_{\earth}}\right) \left(M_{UCD}\over{M_{p}}\right)^{1/3} AU
\end{equation}
where $R_{p}$ and $M_{p}$ are the mass and radius of the planet. The Galilean moon Io orbits Jupiter at a distance of approximately $4a_{\rm R}$, so we will be searching for systems geometrically similar in some sense to the Jupiter-Io system. Since orbits within a few times the Roche limit are physically very close to the UCD, it is possible for planetary companions to UCDs to have very short periods ($P<1$~d).
The physical constraint that the planet orbit outside of the Roche limit requires that

\begin{equation}
P > 0.01 \times \left[ {R_{p}\over{R_{\earth}}} \right]^{3/2} \left[ {M_{p}\over{M_{J}}} \right]^{-1/2} day
\label{pgreat}
\end{equation}
For an Earth-mass planet orbiting a UCD with mass 100 $M_{\rm{J}}$, the period must be $P>0.19$~d.
For these companions the duration of the transit is correspondingly short ($t_{T}< 1 $hr) and the duty cycle for these systems is high

\begin{equation}
\frac{t_{T}}{P}=\frac{1}{\pi}\sin^{-1}\left[\frac{\left(R_{UCD}+R_{p}\right)}{a}\right] \approx \frac{1}{\pi}\left[\frac{\left(R_{UCD}+R_{p}\right)}{a}\right] 
\end{equation}
for edge-on orbits with $\sin{i}=1$. An Earth-mass companion orbiting a UCD with mass 100 $M_{\rm{J}}$ at $a=0.004$~AU with $\sin{i}\approx1.0$ spends approximately $4\%$ of the time in transit. For this system the transit event lasts approximately 20 minutes. The combination of potentially short periods and a relatively high percentage of time in transit makes feasible the targeted transit search that is required by the scarcity of UCDs. In addition, the geometric probability of observing transits, $\pi \times t_{T}/P$, can also be relatively large.

Owing to partial electron degeneracy in their cores, UCDs up to 100 times the mass of Jupiter and older than $10^8$ years have about the same radius as Jupiter \citep{Burrows2001}. This means that over a wide range of UCD mass, transits by terrestrial planets described by the models of \citet{Valencia2005} will produce detectable flux decrements $\Delta F$

\begin{equation}
\frac{\Delta F}{F} \approx 0.008 \left(\frac{M_{p}}{M_{\earth}}\right)^{0.54} 
\end{equation}
A transit by an Earth-mass planet results in an event with a depth of $\approx7$~mmag and a transit by a $10M_{\earth}$ planet results in a transit depth of $\approx0.03$~mag. If photometry with sufficient precision can be produced, it becomes possible, in principle, to detect Earth-size planets with a targeted transit search. 

\begin{figure}
\plotone{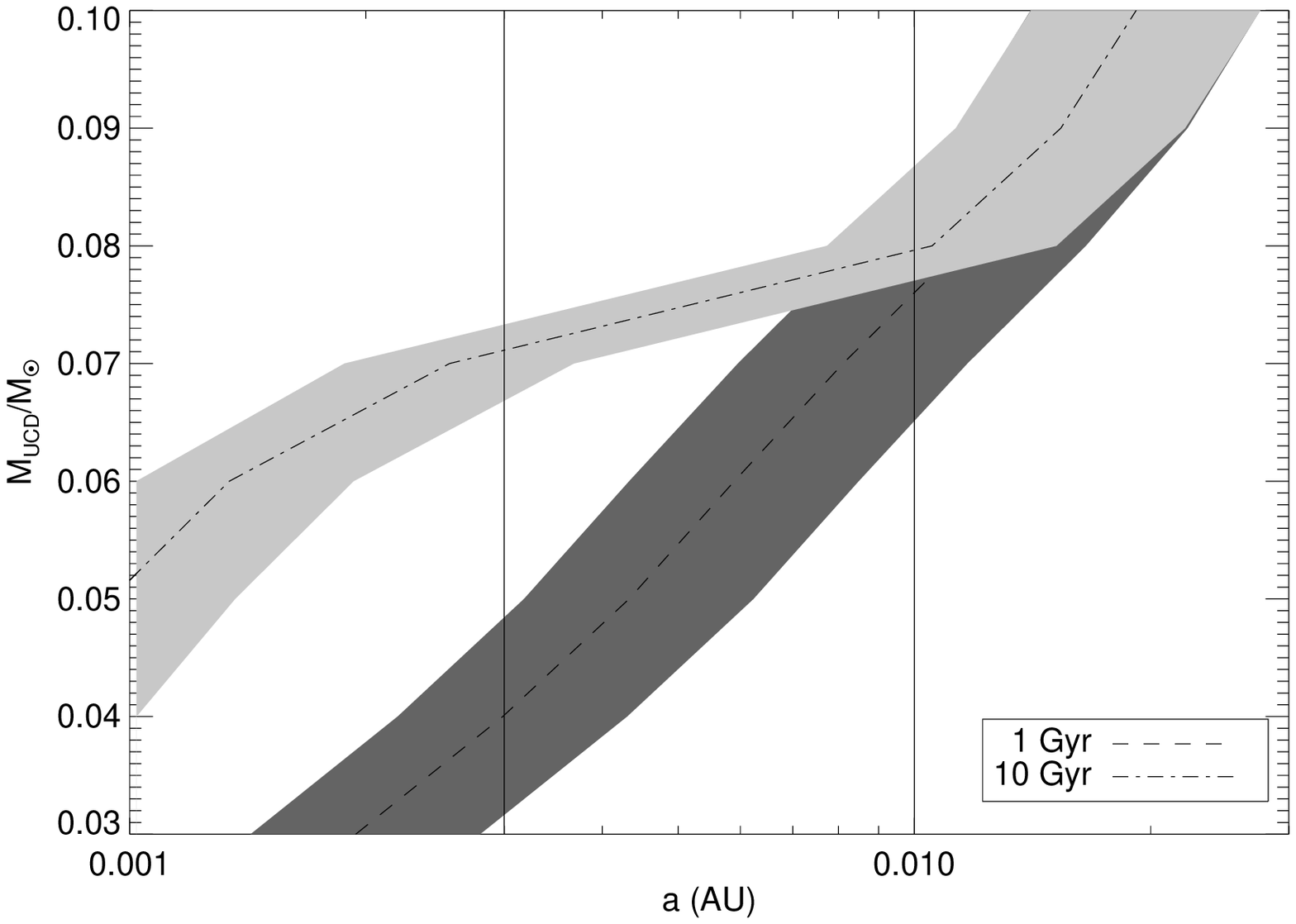}
\caption{Contours of equilibrium temperature $T_{\rm{eq}}=300$~K for planets orbiting at different semi-major axes from UCDs at different ages. These estimates assume a planetary albedo of $A=0.4$ and the UCD models of \citet{Baraffe2003}. The shaded region around each curve represents the temperature range $250<T_{\rm{eq}}<350$~K. The vertical lines indicate the approximate range of semi-major axes to which our targeted transit search could be sensitive.}
\label{tempf}
\end{figure}

\citet{Andreeshchev2004} and \citet{Tarter2007} have pointed out that planets orbiting close to UCDs may spend billions of years in the so-called habitable zone, where temperatures are in the liquid water range. Ignoring the importance of warming due to the greenhouse effect, the equilibrium surface temperature of a planet orbiting close to a UCD at a range of ages can be close to habitable zone temperatures. Since UCDs cool as they age, the orbital
semi-major axis corresponding to liquid water equilibrium temperatures moves with time. Figure \ref{tempf} shows the orbital separation at which the equilibrium temperature, $T_{\rm{ef}}$, is expected to be 250 K to 350 K for a range of UCD masses and ages.

\section{The Peters Automated Infrared Imaging Telescope}
The ability to detect terrestrial planets transiting UCDs is contingent upon the ability to produce precise relative photometry of these objects. 
We carried out observations with the Peters Automated InfRared Imaging TELescope (PAIRITEL) in order to assess the quality of NIR photometry that this system can produce. 
PAIRITEL is a 1.3 meter robotic telescope located at the Fred Lawrence Whipple Observatory on Mt.~Hopkins, Arizona, instrumented with a camera that gathers images in the $J$, $H$, and $K_{S}$ bands simultaneously.  This camera \citep{Skrutskie2006} consists of three $256^2$ pixel NICMOS3 detectors each imaging an $8.5\arcmin$   square field of view at a plate scale of $2\arcsec$ per pixel. This camera was used to conduct the southern portion of the 2-Micron All Sky Survey (2MASS) and the telescope to which it is now attached was utilized for the northern portion.  As a fully robotic observatory, PAIRITEL relies  on a sophisticated software system \citep{Bloom2005} that manages the telescope, camera, and observing plan, monitors the weather, and reduces data as it is gathered. PAIRITEL is able to rapidly respond to Gamma Ray Bursts within two minutes \citep{Blake2005, Yost2005, Perley2007} and, in addition to GRBs and UCDs, has been used to observe X-Ray Bursters \citep{Steeghs2005}, supernovae \citep{Tominaga2005,Friedman2006,Kocevski2007,Wood2008}, Kuiper Belt Objects,  microlensing events \citep{Gaudi2007}, eclipsing binary stars \citep{Blake2008a}, and many other types of targets.  As the largest robotic NIR observatory in the world, PAIRITEL provides unique capabilities for monitoring the time-variable infrared sky.

PAIRITEL is operated in a fixed observing mode in which 7.8s double-correlated ``images'' are created from the difference of a 7.851s and a 51ms integration taken in rapid succession. The telescope is dithered between sets of exposures. This technique, common in NIR imaging, allows for an estimation of the sky and also minimizes the impact of the numerous dead pixels in the NICMOS3 arrays. The dither positions are randomly generated so as to fill an effective area of $10\arcmin$  square  by making many small offsets. Typically, three images are taken at each dither position and a 30 minute exposure may consist of 50 unique dither positions. The efficiency of these observations is up to 70\%, meaning that approximately 21 minutes of integration is gathered during a 30 minute observation.

The automated reduction of PAIRITEL data follows standard NIR data reduction techniques.   The observed number of counts at each pixel, $I_{x,y}$, is the sum of several sources of photoelectrons

\begin{equation}
I_{x,y}=B_{x,y}+D_{x,y}+F_{x,y}\times \left[S_{x,y}+O_{x,y}\right]
\end{equation}

\noindent where $B_{x,y}$ is the detector bias, $D_{x,y}$ is the dark current, $F_{x,y}$ is the flat field response, $S_{x,y}$ is the sky (which may include thermal emission from the telescope and dewar), and $O_{x,y}$ represents the flux from objects such as stars and galaxies, which is what we seek to measure.

For each image an estimate of the bias and sky brightness, called a ``skark'',  is made by taking a median of all the other images acquired within a certain time window. The subtraction of a 51ms exposure from a 7.851s exposure leaves a residual pattern (called shading) that is effectively the bias for the NICMOS3 detector. For a sufficiently large number of images, the process of median combination effectively rejects stars and accurately  reproduces the sky and bias. Assuming that the bias and sky are constant over the time window of the skark, a median of a set of $N$ dithered images results in an estimate of the bias and sky, $SK_{x,y}$

\begin{equation}
SK_{x,y}= Median\left(
{I^{1}_{x,y}}
,...,{{I^{N}_{x,y}}}\right) 
\approx 
{{D_{x,y}}}+B_{x,y}+S_{x,y} \times F_{x,y}
\end{equation}
The dark current in the PAIRITEL detectors is negligible and can be ignored here (ie. $D_{x,y}\approx0$). During the analysis a bad pixel mask is applied to always exclude pixels with highly deviant statistical properties. Particularly in the $K_{s}$ band, the number of bad pixels is large and approaches $1\%$ of the total pixels. 

Bright stars are identified and masked out prior to median combination of an image stack. Experience has shown that stars can effectively be removed and an accurate estimate of the bias and sky brightness can be made using a window of approximately two minutes. Fully reduced images containing counts from the objects of interest are created by first subtracting off the skark image and then dividing by the flat field 

\begin{equation}
O_{x,y} \approx {\frac{I_{x,y}-SK_{x,y}}{F_{x,y}}}
\end{equation}

\noindent We made a master set of PAIRITEL flats from observations of the night sky. As the brightness of the sky decreases (or increases), presumably at the same rate at each pixel on the detectors, the time-dependent change of a pixel value reflects the relative sensitivity of that pixel

\begin{equation}
F_{x,y} \approx  
{\frac{dI_{x,y}}{dS_{x,y}}}
\end{equation}

\noindent An accurate estimate of the pixel response $F_{x,y}$ requires a large range of values of $S$. We created a single master flat in each band for the entire observing campaign by making use of the variations in the sky brightness over all of the observations. Tests on flats created in this way using sets of data from consecutive weeks show differences of $<1\%$. All of these automatic data reduction procedures were carried out 
using codes developed in Python. 

We photometered the fully reduced images following a procedure similar to that described in \cite{Bailer-Jones2003} and \citet{Bailer-Jones2001}. For each UCD a set of several comparison stars within the 8.5$\arcmin$ field of view was chosen in order to create differential magnitudes that remove first order effects due to changes in airmass or transparency. We selected the comparison stars from the 2MASS catalog. We made several cuts at this stage in order to cull bad data. The raw flux of the target UCD was used to discriminate against poor observing conditions by rejecting data where the raw flux was $3\sigma$, typically $>1$ mag, below the rest of the season's observations of that target. Images where a target or comparison falls within 3 pixels of a known bad pixel were also excluded. While we chose the UCDs and comparison stars so as not to saturate the PAIRITEL detectors, bright sky conditions (particularly in K$_{s}$) can lead to non-linearity and saturation. We discarded observations where the maximum pixel value for the UCD or comparison stars exceeded $35000$ ADU.

We measured the flux of the comparison stars and UCD in each image using aperture photometry.  An optimal aperture radius of 3 pixels was empirically determined and used with a sky annulus of radii 6 and 20 pixels. We compared the measured flux of the UCD to the average measured flux of the ensemble of comparison stars to measure a differential magnitude $\Delta m$ in each band 

\begin{equation}
\Delta m=-2.5\log\left({F_{UCD}}\right)+2.5\log\left({{\frac{1}{N}}\sum^{N}_{i=1} \left(F_{C_{i}}/\left<F_{C_{i}}\right>\right) }   \right)
\end{equation}

\noindent where $F_{C_{i}}/\left<F_{C_{i}}\right>$ represents the measured flux of the $i^{th}$ comparison star relative to the mean of all of the other observations of that star

While the individual 7.8s images have reasonable S/N values of up to 100 for the UCDs, we wish to achieve higher precision by binning the observations. We created binned differential magnitudes for each band using a bin size of 300s, or about 25 individual images. The mean of the individual flux measurements in the bin was iteratively calculated after rejecting any measurement that was more than $3\sigma$ from the mean. Working with binned data lets us estimate the errors for each bin using the standard error of the mean of the $N$ individual measurements in the bin after 3$\sigma$ outlier rejection

\begin{equation}
\sigma \left(\Delta m_{N}\right) =N^{-1/2} \sqrt{\frac{1}{N-1} \sum^{N}_{i=1}\left(\Delta m_{i}-\left<{\Delta m}\right>\right)^2}
\label{error}
\end{equation}
In practice we find that the $N$ observations in the bin are not in fact independent measurements as the above equation assumes. The data demonstrate a strong autocorrelation for lags of up to two images. This is due to the dithering pattern employed, where three consecutive individual images are taken at each unique dither position. The result of this significant ``red noise'' is that the achieved photometric precision does not improve as $N^{-1/2}$. The estimate of the error for each  magnitude is taken to be $\sqrt{3}$ times the quantity in Equation \ref{error}. In the limit of negligible limb darkening in the NIR, the signal from a transiting  planet is achromatic. We combine the measurements from the three bands so as to increase the S/N and also mitigate any systematic effects that may be inherent to each band. A simple weighted average was taken of the simultaneous measurements within a five minute time window to calculate a composite magnitude $\Delta m_{JHK}$, represented by the symbol $\Psi$
\begin{equation}
\Delta m_{JHK} = \Psi=\frac{\sum_{j=J,H,K}\left(\Delta m_{j}/ \sigma_{j}^2\right)}{\sum_{j=J,H,K}\left(1/ \sigma_{j}\right)}
\end{equation}
and an estimated error, $\sigma(\Psi)$

\begin{equation}
\sigma \left(\Psi\right)=\left[ {\sum_{j=J,H,K}\left(1/ \sigma_{j}^2\right)}\right]^{-1/2}
\end{equation}

\section{Systematic Errors in NIR Photometry}
There are several pitfalls specific to differential photometry in the NIR that pose potential problems for precise measurements of UCDs. While the techniques of differential photometry in the optical are well developed and have produced impressive results, e.g. \citet{Hartman2005}, the detectors and sky conditions pose unique challenges when working at longer wavelengths, particularly with under-sampled data. 

\subsection{Second-Order Extinction}
In an analysis of the limitations of NIR photometry of UCDs,
\citet{Bailer-Jones2003} discuss the role of extinction by the Earth's
atmosphere in differential photometry. When carrying out such
differential measurements, the assumption is usually made that, to first order,
the amount of extinction is the same for the target and comparison
stars. In the case of UCDs, where the comparison stars are likely to
have significantly bluer and smoother spectra than the targets, this assumption may
not be valid. As the atmosphere above the observatory changes with
time and extinction changes in magnitude and spectral dependence,
the UCD and comparison star are likely to experience different amounts
of extinction relative to each other. These second-order extinction
effects can be potentially large in the NIR where molecular absorption
features dominate the spectrum of the Earth's atmosphere. To further
complicate the issue, these molecular features are known to vary
rapidly, on time scales of minutes. \citet{Bailer-Jones2003} point out that, in general,
these second-order effects are impossible to identify during
otherwise ideal observing conditions and are attributed to
changes in the quantity of water vapor in the atmosphere. This
pernicious effect may in fact represent a practical limit to the
achievable accuracy in ground-based NIR differential photometry from the ground.

To obtain a better idea of the magnitude of the problem, we 
simulated observations of UCDs with a range of comparison stars and
observing conditions. This simulation was based on a detailed model of the atmosphere
at Mt.~Hopkins, high resolution spectra of UCDs and comparison stars,
and the sensitivity of our instrument. The spectral response of the
PAIRITEL system is well known thanks to the detailed analysis done by
the 2MASS team \citep{Cutri2000}.  We used the spectral catalog
of \citet{Pickles1985} as a set of comparison stars. A set of optical
and NIR spectra of UCDs, described by \citet{Cushing2005}, was kindly
provided by that author. Detailed models of the atmospheric
transmission at the Mt.~Hopkins site under a range of observing
conditions were furnished by K. Jucks. These models, calculated using
the the radiative transfer code described in \citet{DesMarias2002}, include molecular
and atomic species as well as aerosols typical of a continental
site. We used a grid of models with Precipitable Water Vapor (PWV) of 1 mm, 3 mm, and 6 mm
and airmasses of 1.0, 1.5, and 2.0. The UCD and
comparison star spectra were first interpolated to the resolution of the
atmospheric model, then convolved with the atmospheric transmission
and telescope response. We integrated the convolved UCD and comparison spectra  to produce an estimate of the differential flux that would
be observed through each of the nine atmospheric models.  As predicted,
we found second-order extinction effects to produce changes in the
flux of the UCD relative to the comparison star with increasing PWV at constant airmass. We calculated the
size of the second-order extinction effect for
comparison star spectral types from B to M and UCDs of type M7 to L4. The results of the simulations are expressed in terms of the
prediction of the \textit{observed} flux differential between the UCD and comparison
star relative to the case of airmass of 1.0 and PWV of 1.0mm. The
relation between the differential magnitude and PWV was found to be nearly
linear so the quantity $d\rm{mag}/dPWV$ gives a reasonable estimate of the observed change in
the differential brightness of the UCD as the PWV increases. We found the $d\rm{mag}/dPWV$
values to be relatively insensitive to airmass. 

We found the importance of the second-order
effects to vary with observing band. In the 2MASS $J$ band, second-order
extinction effects can be relatively large, while in $H$ and $K_{s}$
they are small compared to other sources of noise.  For example, for
an L0.5 UCD with a K0 III comparison star the second-order effect is
estimated to be $-0.0048$~mag/mm in $J$ band, meaning that the UCD appears brighter as the PWV increases. Since the average PWV at sites like Mt.~Hopkins
can easily change by several mm on seasonal time scales or rapidly during poor conditions, the second-order extinction effect
could be as large as a few percent. The red edge of the $J$ band
overlaps with significant atmospheric absorption features, so it is more
susceptible to these second-order effects. This effect has been noted by \citet{Cohen2003} who calculated that the throughput of the 2MASS $J$ band may vary by up to 20\% throughout the year as the average PWV changes. On the other hand, the $H$
and $K_{s}$ bands avoid such atmospheric features and are therefore
more immune. Our calculations
indicate that in $H$ and $K_{S}$ the second-order extinction effect is
likely $<0.3\%$ over the range of reasonable observing conditions. Results of the $J$-band simulations can be found in Table 1. In the 2MASS $J$
band it is possible to produce both short-duration and seasonal dimming or brightening
of $\approx1\%$ in differential photometry of UCDs. In the future it may be possible to monitor the PWV in order to 
identify photometric features which could be due to this effect. We also note that the composite magnitude, $\Psi$, used in our analysis may help to mitigate the effects of PWV since $d\rm{mag}/dPWV$ is found to behave differently in each of the three PAIRITEL bands.

\subsection{Sub-pixel Structure}
The NICMOS3 detectors utilized in the PAIRITEL camera are known to
have response variations at the sub-pixel level. These variations, sometimes called intra-pixel response, are due
to the process by which the chips are manufactured and cause parts of an individual
pixel to be more sensitive to incoming photons than others. The
PAIRITEL PSF is typically under-sampled, with more than 40\% of a
star's total light falling on the single central pixel. Thus, the measured stellar flux could depend on the
precise location of the PSF within the central pixel. This type of effect
becomes small when the starlight is spread out over many pixels, but
when the PSF is very narrow, such as with the NICMOS camera on HST,
large photometric errors can be induced. \citet{Lauer1999} studied
photometry of point sources with the HST NICMOS instrument and found
that in the $J$ band, where the PSF is narrowest, sub-pixel effects can
result in large errors in photometry. In the case of the HST
instrument, the pixels were found to be most sensitive in the center
with a star falling in the center of a pixel appearing nearly 40\%
brighter than those falling near the corner of a pixel. 

We carried out an observational test to estimate the size of the
sub-pixel QE variation effect. The minor planet Aline was observed
while keeping the telescope fixed and not dithering. Aline was
relatively bright ($J\approx12.2$) and moving at a rate of approximately
0.2\arcsec/min across a rich star field when observed for 1.5 hours
by PAIRITEL. Aline has a known photometric period of 12h
\citep{Denchev2000}, much longer than the duration of our
observations. The large number of comparison stars allowed for an
accurate determination of the photometric zero-points of the
individual 7.8s images as well as an accurate astrometric solution for
each image. Just as the sub-pixel QE variations can induce errors in
photometry of narrow PSFs, systematic errors in centroids could also be
induced, resulting in further photometric errors when small apertures are used. The astrometric
solution for each image relative to the ensemble of 2MASS catalog
stars is essentially free from such effects, allowing an estimate of
the centroid of Aline at the Julian Date of each image to better than
0.1 pixel. We also used this data to evaluate the accuracy with which centroids of point sources could be determined. Using a simple center-of-light centroiding method, centroids of individual stars in 7.8s exposures 
were recovered with $\sigma(J)=0.17$, $\sigma(H)=0.25$, and $\sigma(K_{s})=0.24$ pixels. 

Following \citet{Lauer1999}, we tested a model of the
observed flux decreasing as the centroid of the PSF moves
away from the center of the pixel. From this initial test
it was clear that differential flux and sub-pixel radial distance were
correlated, particularly in the $H$ and $K_{s}$ bands. The Spearman Rank test indicated that the $H$ band fluxes of Aline were correlated with both the actual sub-pixel radial position, as determined from the global astrometric fit to the frame, and the position determined by centroiding at a significance of more than $10\sigma$. The observations of Aline involved relatively few pixels, in all of the observations the centroid of Aline landed on one of approximately 80 unique pixels, and only tested the sub-pixel response at a given seeing.

\begin{figure}
\plotone{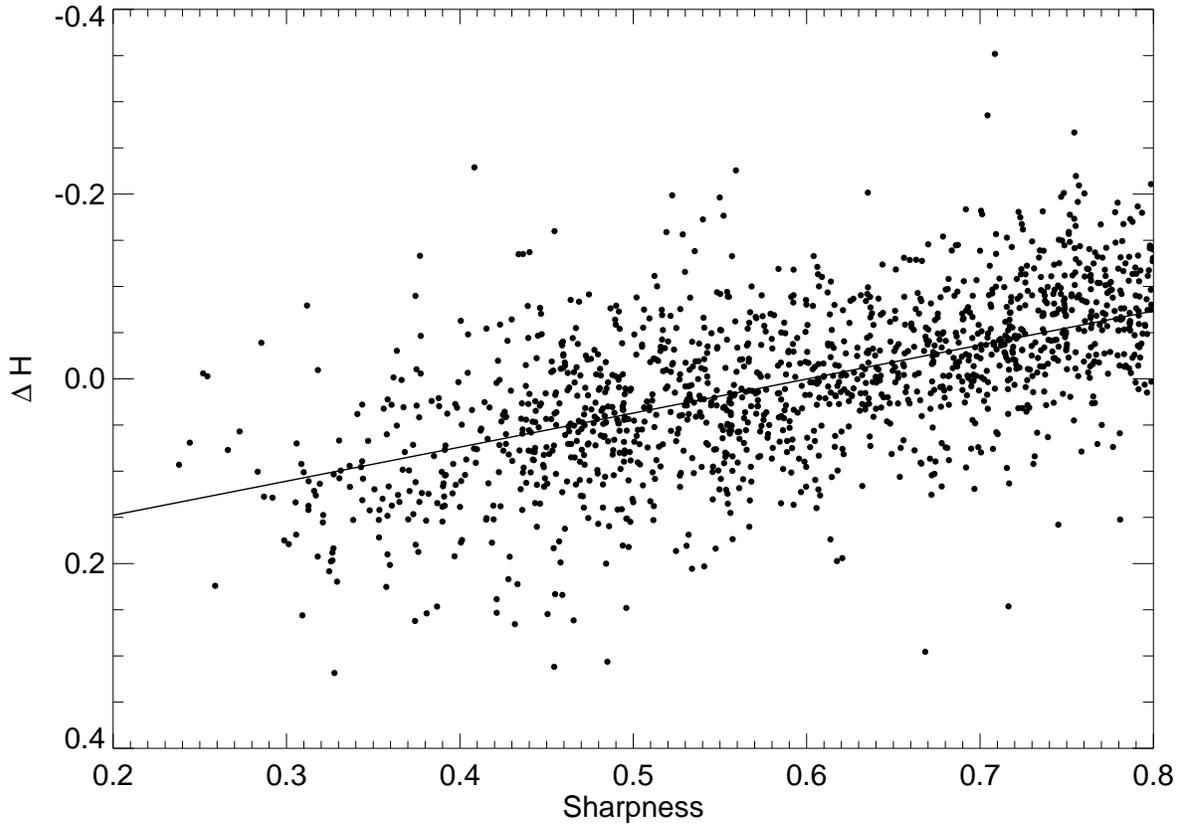}
\caption{Sharpness compared to measured differential magnitude from one night of $H$-band observations. Observations where the UCD is sharper, with relatively more light falling on a single central pixel, result in a systematic overestimation of the brightness of the target. }
\label{hprf}
\end{figure}

In order to avoid the problems of the degeneracy of the center-of-light centroid and the sub-pixel structure, a slightly different approach was taken 
for correcting the UCD observations for this effect. Instead of radial distance from the center of the pixel, the ``sharpness''  of the PSF, defined as the 
ratio of the flux in the central pixel to the total flux, was compared to the measured differential magnitude. This was done for each epoch (30 minute set of observations) allowing for different sub-pixel corrections for different seeing. The average sharpness of the data varied between 0.3 and 0.6 during our observations, so the size of the correction varied significantly. For each epoch we fit the differential magnitude of the UCD as a function of sharpness with a second-order polynomial and we applied the resulting corrections to the differential measurements described in Section 3. An example of an $H$-band sharpness correction is shown in Figure \ref{hprf}.  During a typical set of observations, the size of the sharpness correction can be large, up to 0.1 mag in $H$ band, but it is important to note that stars are statistically unlikely to fall very near the center or very near the edge of a pixel where the sharpness is largest or smallest. Based on the actual histogram of measured stellar centroids in the $H$-band observations and the experiences with the Aline data, the sub-pixel response should contribute approximately 0.013 mag to the observed scatter in the photometry of the individual 7.8s exposures. This agrees well with the realized improvement of 0.014 mag in the scatter of the $H$-band differential magnitudes after application of the sharpness corrections.

\section{Observations and Analysis}

\begin{figure}
\plotone{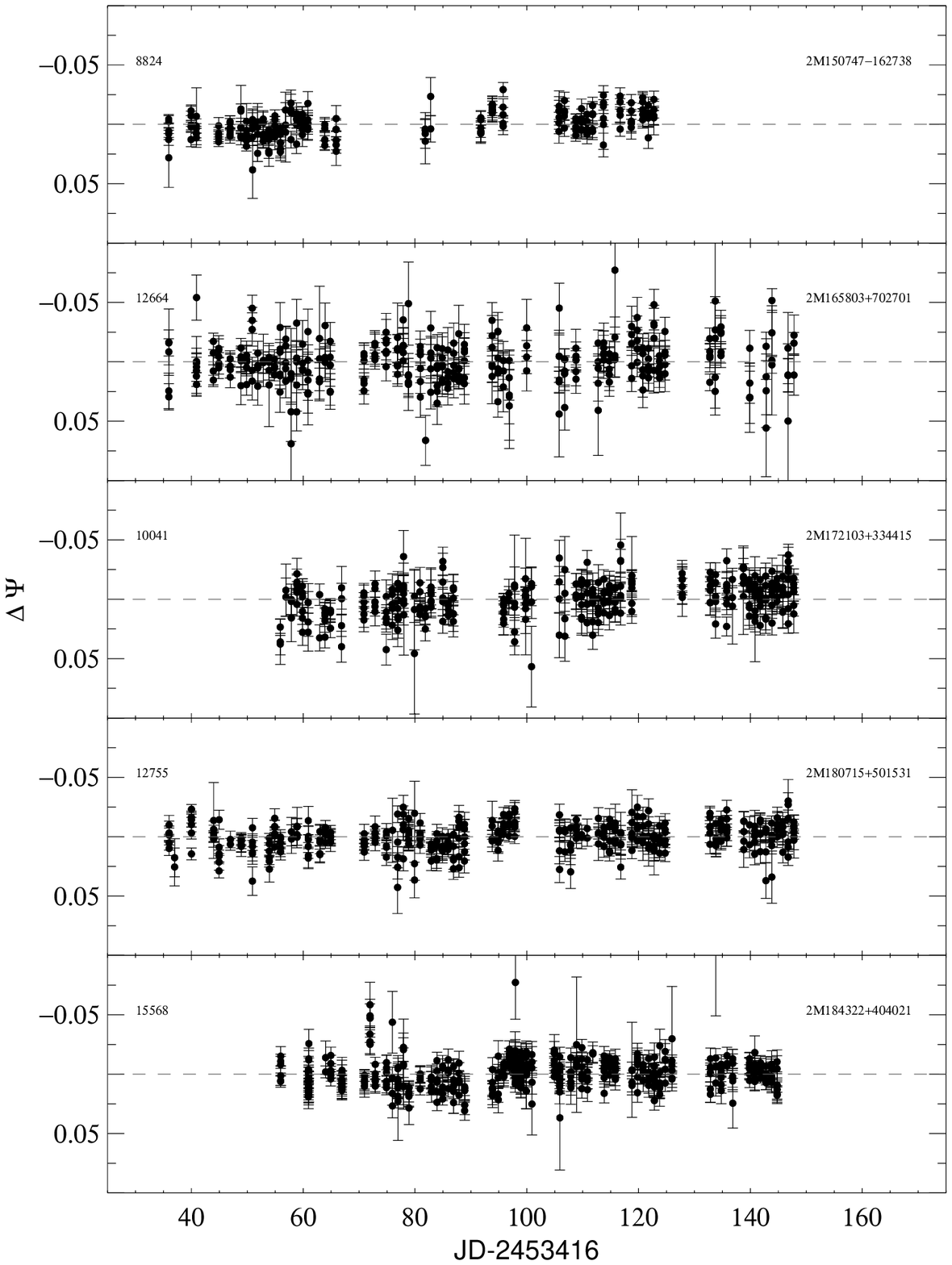}
\caption{Composite J,H,K$_{s}$ ($\Psi$) magnitude light curves of five UCDs observed during the PAIRITEL commissioning phase. The number in the upper left of each panel is the total number of 7.8s observations of each object. Each data point represents a 300 s bin. }
\label{panel1}
\end{figure}

\begin{figure}
\plotone{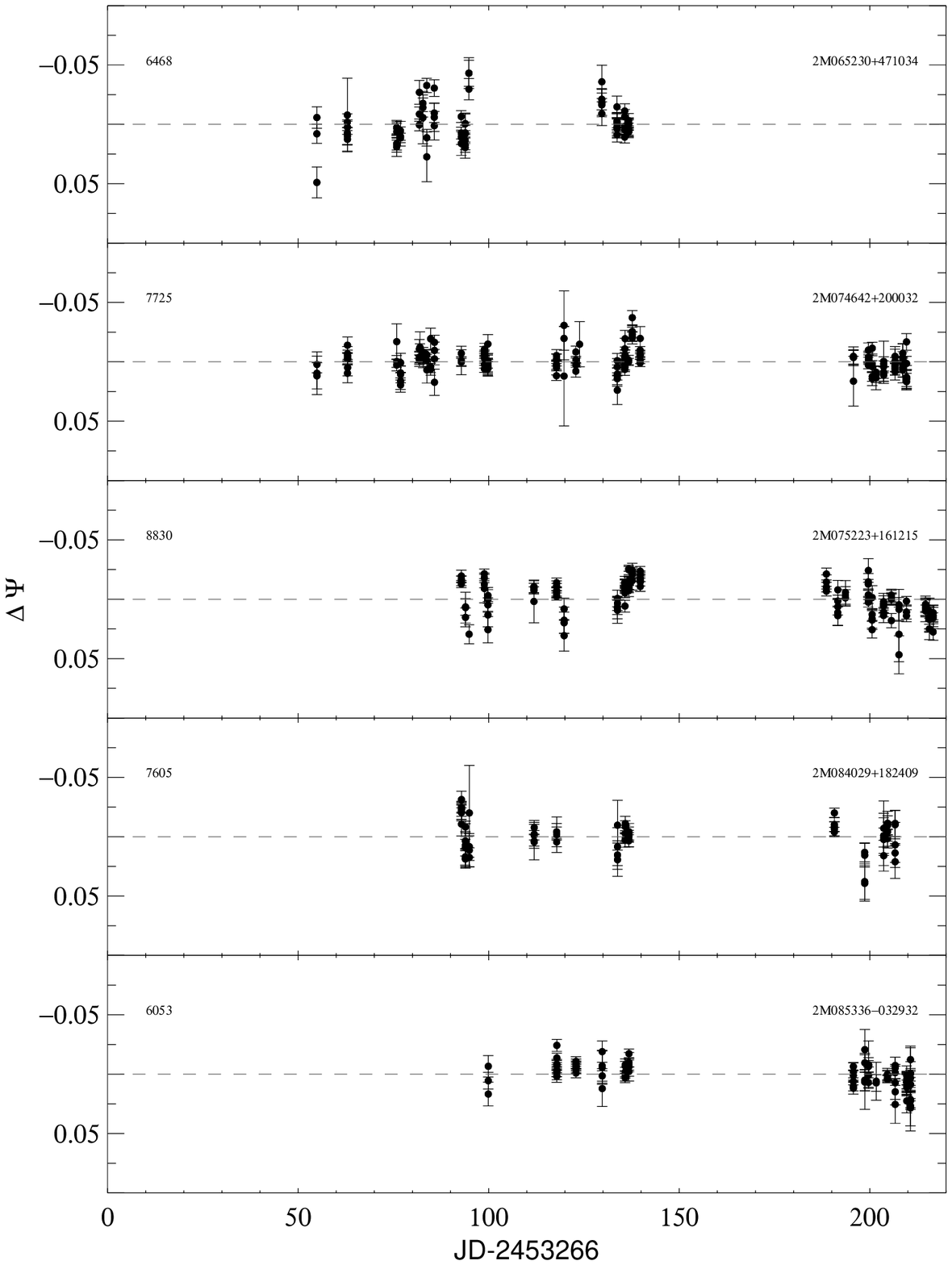}
\caption{Composite J,H,K$_{s}$ ($\Psi$) magnitude light curves of five UCDs observed during the PAIRITEL commissioning phase. The number in the upper left of each panel is the total number of 7.8s observations of each object. Each data point represents a 300s bin.}
\label{panel2}
\end{figure}

We present results from observations of 13 UCDs carried out during the commissioning phase of PAIRITEL. The light curves of ten of the UCDs are shown in Figures \ref{panel1} and \ref{panel2}. When possible, a single 30--40 minute series of exposures was taken each night or every other night for each object.  An overview of the observations of these targets is given in Table 2. The quality of the composite photometry can be characterized in terms of short-term stability and stability over the duration of the observing campaign. The data on short time scales are well described by the photometric errors estimated using the methods described in Section 3. Unfortunately, these errors are at least a factor of  2 to 3 above the expected errors from photon noise alone. The sub-pixel effects discussed in Section 4 are likely an important source of this extra, correlated noise. An analysis of all of the possible sources of this excess noise is beyond the scope of this work. For each of the UCDs we found the composite measurements $\Psi$ combined into 300~s bins to have a scatter of $1-2\%$ over the duration of the observations. The combination of the data from the three observing bands yields significant improvements in the long term stability of the photometry, increases the S/N, and mitigates systematic effects like those discussed in Section 4. Since the transit event is expected to be achromatic, combining the measurements from the three bands should not inhibit our ability to measure transit properties. 

These data are very useful for characterizing the variability of UCDs and a detailed analysis of the variability of the targets in our sample will be presented in a future work. There is a large body of literature on the variability of UCDs and how observed variability may relate to rotation and dynamic dust and cloud features in their atmospheres (\citealt{Gelino2002}, \citealt{Bailer-Jones2003}, \citealt{Koen2005}). The PAIRITEL observations, with simultaneous measurements in $J$, $H$, and $K_{s}$ may prove particularly useful for identifying color-dependant variability as predicted by some models of UCDs \citep{Bailer-Jones2008}. At the same time, combining the simultaneous measurements into the composite magnitude $\Psi$ may help to mitigate the effect of intrinsic variability if the variability is uncorrelated or anti-correlated between the bands. Here, we are mainly concerned with variability that may inhibit the detection of transit events. We find little evidence for variability with an amplitude greater than 0.02 mag in any of our light curves and find that the data are generally well-described by our photometric error estimates as outlined in Section 3. As a simple statistical test we calculate the reduced Chi-squared, $\chi_{\nu}^2$, for each light curve (300s bins) assuming the null hypothesis of no photometric variability. These results, along with the RMS of each light curve, are given in Table 3. In all but two cases we find $\chi_{\nu}^2<4$, indicating that we observe little variability in excess of the typical photometric errors of 0.01 mag.  The UCD with the largest values of $\chi_{\nu}^2$, 2M0752+1612, is also the brightest target in our sample, indicating that non-linear effects in the detector may become important for objects with $J<10.8$.

To assess our ability to recover transit signals from the PAIRITEL data, we carried out a Monte-Carlo simulation. The methods for searching time series data for box-like transits are well developed. The Box Least Squares (BLS) method developed by \citet{Kovacs2002} has been used to successfully find transiting planets and is in general use by the transit search community. Since each individual UCD has relatively few ($\approx 300$ binned measurements) we decided to combine all of the observations of five of our targets for use in the simulations. We combined the observations of five of the UCD targets (2M1507$-$1627, 2M1658+7027, 2M1721+3344,2M1807+5015,2M1843+4040) in order to create a hypothetical set of observations that would simulate an intensive observing campaign on one target. Using this hypothetical set of observations, we inserted transits of different durations, amplitudes, periods, and phases into the data and we used the BLS method to try to recover the inserted transit event. We determined the transit parameters following the work of \citet{Mallen2003} and the scaling relations in Section 2. For these simulations we used a fixed UCD mass of $100M_{\rm J}$. We considered Super-Earth planets between 3 and $10M_{\earth}$  described by the models of \citet{Valencia2005}. The simulation was parameterized in terms of the Roche limit of the planet-UCD system and we considered orbits between 1 and $5R_{\rm R}$ (approximately 0.003 to 0.01 AU).

\begin{figure}
\plotone{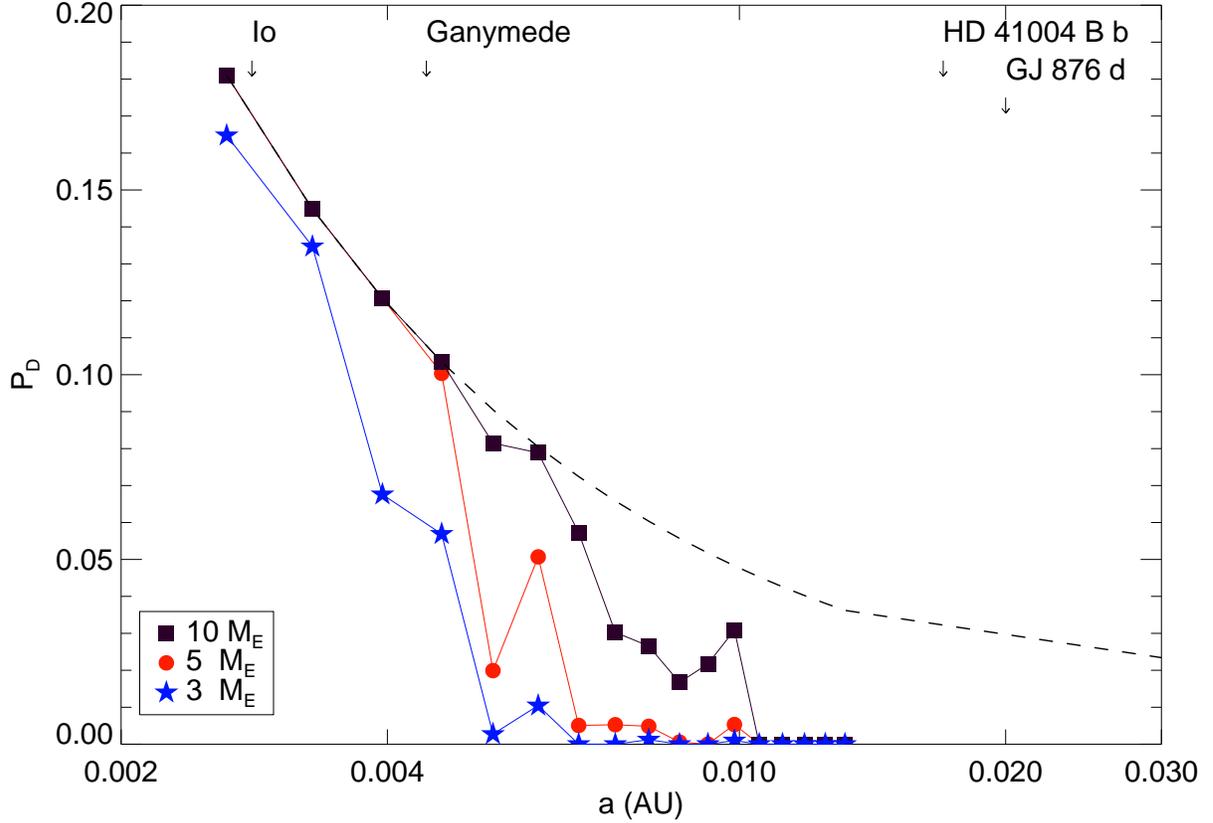}
\caption{Results of the simulation of planet detection efficiency using the hypothetical data from an intensive campaign to observe a target for 130 hours over three months. The dashed line is the geometric expectation, $(R_{UCD}+R_{P})/a$. The orbital separations of two moons of Jupiter, Io and Ganymede, as well as two known extrasolar planets, are shown. The orbits considered in the simulations extend from the Roche limit at $\approx 0.003$ AU. The radius of Jupiter is $R_{\rm J}$=0.0004 AU, so the distance between a UCD and a planet orbiting at the Roche limit is still several times the radius of the UCD.}
\label{sim}
\end{figure}

The BLS algorithm was applied to the hypothetical data by searching for events with frequencies in the range 0.5 to $10~\rm{d}^{-1}$ with a frequency resolution of $0.001~\rm{d}^{-1}$ and transits with fractional durations between 0.01 and 0.15 of a period. After visual inspection of a set of 100 simulated transit light curves and the corresponding BLS spectra, we created an empirical set of criteria for successful detection of a transit in a simulated light curve.
The recovery of a transit was considered successful if a peak in the BLS power spectrum was within 0.002 d$^{-1}$ of the correct (inserted) period and at least 7 $\sigma$ above the average power level in the BLS power spectrum. The results of these simulations are shown in Figure \ref{sim}. In this representation,  the geometric detection probability ($\frac{R_{UCD}+R_{P}}{a}$) has been factored in so these probabilities represent the likelihood that a planet in a given orbital configuration and random inclination would have been detected around a single star during a season of PAIRITEL observations. These probabilities implicitly include the noise peculiarities of the actual data and losses due to poor observing conditions or instrumental malfunctions. We find that the terrestrial planet detection efficiency for single targets is above $5\%$ for orbital separations $a<0.004$ AU for planets larger than $3M_{\earth}$. With 40 minute binning the RMS of the observations of five of the UCD targets drops to 7 mmag or less. The expected signal from a $1M_{\earth}$ companion transiting a UCD
is approximately 8 mmag. A simulation with 40 minute binning indicated that we currently have no sensitivity to planets with $M<1$M$_{\earth}$. This suggests that if the sources of excess noise in the PAIRITEL observations could be identified and alleviated, and 7 mmag precision could be achieved with smaller bins in time, it might become possible to detect Earth-mass planets in the habitable zones of their hosts.

While the $\approx0.03$ mag signal from a $10M_{\earth}$ transiting companion is readily detectable in our simulations, the sensitivity to Earth-mass planets will depend critically on the noise properties of the NIR photometry. In Section 3 we discussed the correlated or ``red noise'' in the PAIRITEL photometry on short ($< 1$ minute) timescales, which limits our photomeric precision to well above the expected noise from photons alone. For the detection of subtle transit signals, \citet{pont2006} highlighted the importance of red noise for the estimation of detection efficiency, particularly the red noise on a timescale comparable to the transit event. For short-period, terrestrial companions to UCDS the expected transit duration is $t_{T} \approx 30$ minutes. Following \citet{pont2006} we estimate the 
red noise in the hypothetical data used in the Monte-Carlo simulations on the transit timescale to be $\sigma_{r}=6.9$ mmag. The results of our simulations are generally consistent with findings in \citet{pont2006} who suggest the detection statistic  

\begin{equation}
S_{r}^{2}=\frac{d^{2}}{n^{-1}\sigma_{w}^2+N_{tr}^{-1}\sigma_{r}^2}
\end{equation}
where $d$ is the transit depth, $n$ is the total number of points in transit, and $\sigma_{w}$ is the white noise.
For 3,5, and 10 $M_{\earth}$ planets orbiting at $a=0.004$ AU we estimate that the detection statistics $S_{r}$ are 6.1, 8.1, and 11.3, respectively. The reliable detection of transit signals generally requires $S_{r}>8$. In our simulations we find that the detection of $3M_{\earth}$ companions at this orbital separation is marginal as is expected based on the $S_{r}$ criteria. The detection of an Earth-mass companion with $d=8$ mmag will require significant improvements in $\sigma_{w}$, which we feel are likely achievable, but also improvements in $\sigma_{r}$. Improving the photometric precision from $\sigma_{w}=0.01$ to $\sigma_{w}=0.005$ mag will only improve the expected detection statistic for an Earth-mass companion from $S_{r}=3.2$ to $S_{r}=3.6$. If both $\sigma_{r}$ and $\sigma_{w}$ can be reduced to the level of 3 mmag, then detection of Earth-mass planets will likely become feasible.

Placing a useful limit on the rate of super-Earth companions in orbits with $a<4R_{\rm R}$ will require increasing the phase coverage of the individual targets in the PAIRITEL sample by a factor of approximately five and increasing the sample size to 60 or more objects.  With a sample of this size we expect to detect at least one transiting planet ($95\%$ confidence) if every UCD target has a close-in ($a<0.01 AU$), $M>5M_{\earth}$ companion. The hypothetical set of observations used in the transit detection efficiency simulations, comprised of the combined data from five UCDs, represents about 1600 observations of five minutes each, or about 130 hours of observing time. Now that the performance of the PAIRITEL system is better optimized, it 
is possible to realize about $60\%$ of available observing time (including losses due to weather), so the 130 hours of observations would require approximately 22 nights to acquire. Over a three month observing campaign it would be possible to observe four targets with the required cadence. At this rate, sufficient observations of a sample of 60 UCDs could be amassed in approximately four years. 

The short periods of close-in companions to UCDs ($P<1$d) allow the possibility of detecting planets with different observing strategies. With telescopes distributed in longitude, allowing for 24 hour observations, it is possible to continuously monitor a single UCD with a telescope network like Las Cumbres Global Observatory\footnote{\url{http://www.lcogt.net}}. By continuously monitoring a single UCD for 48 hours, two transits could be detected for companions with $P<1$~d. Without losses due to poor weather, such a telescope network could detect terrestrial companions orbiting at twice the Roche limit or closer and reveal single transits by planets at four times the Roche limit. A network of telescopes, each able to produce $1\%$ photometry of UCDs, and distributed in both latitude and longitude, would be robust against poor weather and could acquire the necessary observations of the same sample of 60 targets in approximately four months. 

Candidates discovered by a UCD transit survey will require radial velocity (RV) measurements in order to both confirm the planet
and measure its mass. Since UCDs are very red and usually faint, the RV measurements will pose a significant challenge. Work by \citet{Blake2007a, Blake2008b} has demonstrated the potential for obtaining RVs with high-resolution NIR spectroscopy. For targets brighter than $K=12.0$, an RV precision of 200 m s$^{-1}$ has been demonstrated. Eventually, it is expected that these observations will produce 50 m s$^{-1}$ RV measurements. Assuming a UCD mass of $100M_{\rm{J}}$, a terrestrial companion with mass $5M_{\earth}$ in an orbit $a<0.01$ AU produces a radial velocity semi-amplitude of $K>  14$ m s$^{-1}$. This RV signal is likely undetectable with current instruments, but a future generation of NIR spectrographs on large telescopes could make this possible. Most transit searches suffer from astronomical false positives, which 
need to be ruled out with precise radial velocities \citep{Charbonneau2004}. In the case of UCD targets, the false positive due to a small star transiting in front of a larger star (the F+M binary case) is very unlikely since the targets selected as UCDs are almost certain not to be larger, main sequence stars. It is also possible to mimic a transit signal through the chance alignment of an eclipsing binary with a background star. While an eclipsing UCD binary would be an exciting discovery in its own right, systems in the period range ($P<1$~d) to which our search would be most sensitive are very rare \citep{Burgasser2007}. Furthermore, the large proper motions observed for many UCDs \citep{Deacon2007} would make the chance alignment with a background star, a star that would have to similar colors to the UCD, short lived.

\section{Conclusions}
We present NIR observations of 13 UCDs that are photometrically stable over several months. We demonstrate that UCDs are viable targets for searches for transiting extrasolar planets and that an intensive observing campaign with a single telescope spanning several years could detect terrestrial companions potentially located in the habitable zones of their UCD hosts. Similar planets could also be found with a campaign spanning less than six months with a world-wide network of of small aperture telescopes (such as envisioned by Las Cumbres Observatory). The current photometric precision ($\approx 1\%$) is sufficient for the detection of companions as small as $3M_{\earth}$ and, with future improvements, it may be possible to detect companions as small as Earth.  A UCD transit search is currently one of the most promising ways to search from the ground for terrestrial planets in the habitable zones of their hosts. Since bright UCDs are relatively rare, such a search needs to be carried out one object at a time. We have simulated our transit detection efficiency and found that with a long time baseline our observing strategy can effectively recover events with short periods. To obtain the required phase coverage to detect planets in short period orbits it is necessary to observe each target over many months. We estimate that a four year program to observe 60 UCDs could place a significant upper limit on close-in, super-Earth to Neptune size companions to UCDs. While analysis of current transit searches has led to the conclusion that fewer than 1 in 300 main sequence stars has a close-in massive companion \citep{Gould2006a}, we know virtually nothing about the occurrence of planetary companions to UCDs or the rate of super-Earth companions to any type of star. Observations of planet-forming disks around UCDs lead us to believe that it is likely possible to form planets around these low mass hosts. Future observations of these small stars, both by searching for transits in the NIR and gathering increasingly precise NIR radial velocity measurements, will hopefully reveal an exciting new class of planets which are susceptible to study by direct imaging and provide important tests of models of terrestrial planets and their formation.

\acknowledgments The authors would like to thank an anonymous referee for thoughtful comments that
helped to improve this manuscript.
It is a pleasure to acknowledge S. Gaudi for his many helpful comments and suggestions. The authors
would also like to thank G. Bakos, D. Charbonneau, D. Finkbeiner, A. Gould, 
J. Hora, M. Wood-Vasey, and C. Stubbs for helpful conversations that
contributed to this work. Thanks also to C. Cushing, K. Jucks,
D. Fabryky, and G. Torres for contributing code and data that
facilitated this work. Finally, it is pleasure to thank D. Spergel for
his support and encouragement in developing the concept of the brown
dwarf transit search. CB thanks NASA's Kepler Mission for partial
support via Cooperative Agreement NCC2-1390 and support from the
Harvard University Origins of Life Initiative. JSB is partially
supported by a Sloan Research Fellowship.

The Peters Automated InfRared Imaging TELescope (PAIRITEL) is operated
by the Smithsonian Astrophysical Observatory (SAO) and was made
possible by a grant from the Harvard University Milton Fund, the
camera loan from the University of Virginia, and the continued support
of the SAO and UC Berkeley. The PAIRITEL project is further supported
by NASA/Swift Guest Investigator Grant NNG06GH50G. We thank
M. Skrutskie for his continued support of the PAIRITEL project, as well as the staff at Fred Lawrence Whipple Observatory. This
research has benefited from the M, L, and T dwarf catalog archived at
DwarfArchives.org and maintained by C. Gelino, D. Kirkpatrick,
and A. Burgasser.

\clearpage

\clearpage

\clearpage

\clearpage

\clearpage

\begin{deluxetable}{lccccccc}
\tabletypesize{\scriptsize} \tablecaption{$J$-band Second-Order
Extinction} \tablewidth{0pt} \tablehead{ \colhead{$UCD$} &
\colhead{$B0III$} & \colhead{$A0III$} & \colhead{$F0III$} &
\colhead{$G0III$} & \colhead{$K0III$} & \colhead{$M4V$} &
\colhead{$G2V$} } \startdata L0.5 & -0.0033 & -0.0034 & -0.0037 & -0.0046
& -0.0048 & -0.0052 & -0.0043\\ L1 & -0.0028 & -0.0028 & -0.0032 & -0.0040 &
-0.0042 & -0.0047 & -0.0038 \\ L2 & -0.0020 & -0.0020 & -0.0024 & -0.0032 &
-0.0034 & -0.0038 & -0.0030 \\ L3 & -0.0014 & -0.0014 & -0.0018 & -0.0027 &
-0.0029 & -0.0033 & -0.0024 \\ M9 & -0.0005 & -0.0005 & -0.0009 &-0.0017 &
-0.0019 & -0.0024 & -0.0015 \\ M7 & -0.0000 & -0.0000 & -0.0004 & -0.0012 &
-0.0014 & -0.0018 & -0.0009 \\ \enddata
\tablecomments{Estimated values of
$d\rm{mag}/dPWV$ for differential photometry of UCDs
with different comparison stars. Units are magnitudes per $mm$ of PWV.}
\label{table1}
\end{deluxetable}

\begin{deluxetable}{lccccccc}
\tabletypesize{\scriptsize} \tablecaption{UCD Targets} \tablewidth{0pt}
\tablehead{ 

\colhead{Object} & \colhead{Sp. Type} & \colhead{$J$} &
\colhead{$H$} & \colhead{$K_{s}$}  &
\colhead{Comps} & \colhead{$T_{max}$} & \colhead{Obs} \\ 

\colhead{} & \colhead{} & \colhead{} & \colhead{} & \colhead{} &  \colhead{} & \colhead{(days)} & \colhead{}}\startdata

2M0045+1634$^{a}$ & L3.5 & 13.06 & 12.06 & 11.37  & 3 & 208.2 & 61\\ 

2M0213+4444 & L1.5 & 13.51 & 12.77 & 12.24 &  30 & 208.3 & 44\\

2M0320$-$0446 & M8 & 13.25 & 12.54 & 12.11 & 7 & 81.9 & 130 \\

2M0652+4710 & L4.5 & 13.55 & 12.38 & 11.69 &  16 & 144.8 & 86\\ 

2M0746+2000 & L0.5 & 11.74 & 11.00 & 10.49 & 15 & 154.7 & 167\\ 

2M0752+1612 & M7 & 10.83 & 10.19 & 9.82 &  15 & 129.8 & 61\\

2M0840+1824 & M6 & 11.05 & 10.40 & 10.05 &  8 & 117.8 & 86\\

2M0853$-$0329 & M9 & 11.185 & 10.47 & 9.97 & 10 & 115.0 & 114 \\

2M1507$-$1627 & L5 & 12.82 & 11.90 & 11.62 & 12 & 86.8 & 204 \\

2M1658+7027 & L1 & 13.31 & 12.54 & 11.92 & 4 & 111.9 & 315 \\

2M1721+3344 & L3 & 13.58 & 12.92 & 12.47 & 12 & 91.9 & 322\\

2M1807+5015 & L1.5 & 12.96 & 12.15 & 11.61  &  7 & 111.8 & 376\\

2M1843+4040 & M8 & 11.30 & 10.67 & 10.27 & 7 & 88.8 & 433\\

 \enddata 
 
 \tablecomments{Spectral type and $J$, $H$, $K_{s}$ magnitudes taken from \citet{Cruz2003},
except where noted. The value of $Comps$ indicates the number of comparison stars used in the differential photometry. The value of $T_{max}$ corresponds to the span of time covered by the first and last observation of each object. The $obs$ value indicates the number of 300s observations of each target.}
\tablenotetext{a}{Data from \citet{Wilson2003}}
\label{table2}
\end{deluxetable}

\begin{deluxetable}{lccc}
\tabletypesize{\scriptsize} \tablecaption{Light Curve Statistics} \tablewidth{0pt}
\tablehead{ 

\colhead{Object}  & \colhead{$RMS_{5}(\psi)$} & \colhead{$RMS_{40}(\psi)$} &
\colhead{$\chi_{\nu}^{2}$}\\ 

\colhead{} &  \colhead{(mmag)} & \colhead{(mmag)} & \colhead{}}
\startdata

2M0045+1634 & 17 & 10 & 0.97\\ 

2M0213+4444 &  17 & 15 & 2.20\\

2M0320$-$0446 & 21 & 14 & 4.57  \\

2M0652+4710 &  14 & 13 & 3.29\\ 

2M0746+2000 &  9 &  6 & 2.47 \\ 

2M0752+1612 & 14  & 13 & 7.37 \\

2M0840+1824 &   14 & 10 & 2.68 \\

2M0853$-$0329 & 10 & 7 & 2.10  \\

2M1507$-$1627 & 11  & 7 & 2.57  \\

2M1658+7027 & 17  & 8 & 1.56 \\

2M1721+3344 & 14 & 8 & 1.78 \\

2M1807+5015 &11 & 6 &  2.10\\

2M1843+4040 &  11 &7&  1.92 \\

 \enddata 
 
 \tablecomments{The quoted RMS$_{5}(\Psi)$ and  RMS$_{40}(\Psi)$ values are based on 5 minute and 40 minute binning. $\chi_{\nu}^{2}$ is the Chi-squared per degree of freedom assuming the null hypothesis of no photometric variability.}
\label{table3}
\end{deluxetable}

\end{document}